# Directly grown monolayer MoS$_2$ on Au foils as efficient hydrogen evolution catalysts

Jianping Shi[1,2*], Donglin Ma[2*], Gao-Feng Han[3], Yu Zhang[1,2], Qingqing Ji[2], Teng Gao[2], Jingyu Sun[2], Cong Li[1], Xing-You Lang[3★], Yanfeng Zhang[1,2★], Zhongfan Liu[2]

[1] Department of Materials Science and Engineering, College of Engineering, Peking University, Beijing 100871, People's Republic of China

[2] Center for Nanochemistry (CNC), Beijing National Laboratory for Molecular Sciences, College of Chemistry and Molecular Engineering, Academy for Advanced Interdisciplinary Studies, Peking University, Beijing 100871, People's Republic of China

[3] Key Laboratory of Automobile Materials (Jilin University), Ministry of Education, School of Materials Science and Engineering, Jilin University, Changchun 130022, People's Republic of China

★ Corresponding Author: yanfengzhang@pku.edu.cn; xylang@jlu.edu.cn

* These authors contributed equally to this work.

**Synthesis of monolayer $MoS_2$ is essential for fulfilling the potential of $MoS_2$ in catalysis, optoelectronics and valleytronics, *etc*. Herein, we report for the first time the scalable growth of high quality, domain size tunable (edge length from ~ 200 nm to 50 μm), strictly monolayer $MoS_2$ on commercially available Au foils, *via* a low pressure chemical vapor deposition method. The nanosized triangular $MoS_2$ flakes on Au foils was proved to be an excellent electrocatalyst for hydrogen evolution reaction (HER), featured by a rather low Tafel slope (61 mV/decade) and a supreme exchange current density (38.1 μA/cm$^2$). The abundant active edge sites and the excellent electron coupling between $MoS_2$ and Au foils account for the extraordinary HER activity. Our work presents a sound proof that strictly monolayer $MoS_2$ assembled on a well selected electrode can manifest comparable or even superior HER property than that of nanoparticles or few-layer $MoS_2$ electrocatalyst.**

The significance of two dimensional (2D) atomic layer thin materials has been fully demonstrated by a series of fascinating performances emerged in graphene.[1,2] However, the zero band gap property of graphene impedes its applications in carbon-based nanoelectronics and optoelectronics.[3] In this regard, transition metal dichalcogenides (TMDCs) with lamellar structures have subsequently drawn great interest due to its sizable band gap with indirect to direct tunability from bulk to one atomic layer,[4-6] and plenty of intriguing performances in electrical,[7] optical,[8] and photovoltaic devices.[9]

Synthesizing uniform monolayer TMDC films with high crystal quality and large domain size should be the premise to fulfill the application potential of TMDC, and a variety of synthesis methods have been developed so far. Top-down micromechanical exfoliation[10] and ionic intercalation[11] were firstly employed to produce atomically thin TMDC films, the production of which was nevertheless of non-uniform thickness and versatile size distributions, thus incompatible with efficient batch fabrication. Recently, several bottom-up methods, such as transition metal sulfurization,[12] decomposition of thiomolybdates,[13] and physical vapor deposition,[14] have been explored to synthesize TMDC films on insulating substrates, but still resulting in uncontrollable film thickness ranging from monolayer to few-layer. Compared with abovementioned routes, chemical vapor deposition (CVD) technique is more simple and efficient in synthesizing monolayer TMDC films[15-20] because of its wide tunability in growth parameters and substrates.[21] Note that, the CVD TMDC films growth on insulating substrates ($SiO_2$,[22] mica,[23] sapphire[24]) has demonstrated their perfect application potentials in electronic and photovoltaic devices.

In particular, theoretical calculations have indicated that the free energy of hydrogen bonding to the sulfur edge of $MoS_2$ (a member of TMDC) is close to that of Pt,[25] which makes $MoS_2$ a potentially low-cost, high-abundance substitute of Pt in electrocatalytic hydrogen evolution reaction (HER).[26-29] To this end, $MoS_2$-based HER electrocatalysts such as amorphous particles,[30] chemically exfoliated nanosheets,[31-33] mesopores,[34] as well as thin films[35-36] have been intensively synthesized in various routes. Although rather high HER activities have been obtained therein, the HER mechanism is still inconclusive due to the complexity of the reaction, probably mediated by a wide distribution of particle size and layers thickness, and the multiple surface texture, *etc.* Consequently, seeking for ideal systems, e.g. 2D monolayer $MoS_2$ nanoislands with tunable sizes, would be helpful in establishing a direct correlation between the catalytic activity and the microscopic structure (namely the amount of edge sites), considering that the HER activity originating from the edges of $MoS_2$ whilst the basal planes are catalytically inert.[25,37] It is worthy of mentioning that ultrahigh vacuum (UHV) deposition and recognition of monolayer $MoS_2$ nanoislands was realized on Au(111) by scanning tunneling microscope (STM).[38] The as-grown sample presented a rather low Tapfel slope of ~ 55-60 mV/decade and a rather high HER activity. However, the fabrication process involved in obtaining $MoS_2$/Au(111) was nevertheless instrument-dependent and cost-ineffective, hence retarding the large-scale production and the practical application of monolayer $MoS_2$.

Herein, we demonstrate, for the first time, the scalable synthesis of monolayer MoS$_2$ on commercially available Au foils *via* a facile low-pressure CVD (LPCVD) method. Notably, the domain size of monolayer MoS$_2$ flakes can be tuned from nanometer to micronmeter by varying the growth temperature or the precursor substrate distance. This is beneficial for the utilization of monolayer MoS$_2$/Au directly in efficient HER and in constructing electronic and optoelectronic devices after a suitable sample transfer process. The domain size, crystal quality, and thickness of as-synthesized MoS$_2$ on Au foils were probed by various characterization techniques. Moreover, the HER performance of as-grown nanosized monolayer MoS$_2$ on Au foils was examined as a function of film coverage or edge density. The mechanism of the enhanced HER activity was investigated for the LPCVD grown MoS$_2$ on Au foils.

Figure 1a schematically illustrates the formation of MoS$_2$ 2D flakes on Au foils, where the MoO$_3$ powder was partially reduced by S to form volatile suboxide species of MoO$_{3-x}$, and then sulfurized into MoS$_2$ on the downstream Au foils with the aid of Ar carrier gas, as described by the reaction:

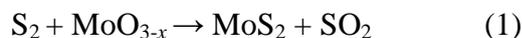

$$S_2 + MoO_{3-x} \rightarrow MoS_2 + SO_2 \qquad (1)$$

The LPCVD setup employed in this work is depicted in Fig. 1b (See Methods for further details of the growth methodology). The X-ray photoemission spectroscopy (XPS) spectra (Fig. 1c and Supplementary Fig. S1) was captured to confirm the formation of MoS$_2$, as characterized by the Mo 3d$_{5/2}$ (at 229.1eV) and Mo 3d$_{3/2}$ (at 232.3eV) peaks associated with Mo$^{4+}$, as well as the S 2p$_{3/2}$ (at 162.1eV) and S 2p$_{1/2}$ (at 163.2eV) peaks assigned to $S^{2-}$, which are all in good agreement with those previously reported for $MoS_2$.[15]

Scanning electron microscope (SEM) (Fig. 1d-g) micrographs were then captured to show the domain size evolution of as-grown $MoS_2$ on Au foils. Notably, after 1 h growth at 530 ℃ (Fig. 1d), triangular $MoS_2$ flakes with a nearly uniform edge lengths of ~ 200 nm can be obtained on Au foils. Upon increasing the growth temperature to 680 ℃ while keeping other growth parameters identical (Fig. 1e and Supplementary Fig. S2), the edge length of triangular $MoS_2$ flake can be enlarging to ~ 55 μm. Further increasing the growth temperature to ~ 750 ℃ leads to the decline in edge length (~ 30 μm) of the triangular domains (Fig. 1g). Intriguingly, the 2D large domain $MoS_2$ flakes (formed at 680 ℃) can be visible by optical microscope (OM) (Fig. 1h). It is worthy of mentioning that, with our LPCVD method, the area of as-grown $MoS_2$ sample is only limited by the size of Au foils, where the photograph in Fig. 1i represents a typical sample from the batch production with a 3 × 3 cm$^2$ coverage area.

Raman spectroscopy measurements were carried out to probe the layer thickness as well as the crystal quality of the as-grown $MoS_2$ samples on Au foils (Fig. 1j). The out-of-plane vibration of S atoms ($A_{1g}$) at ~ 406.9 cm$^{-1}$ and in-plane vibration of Mo and S atoms ($E_{2g}^1$) at ~ 387.0 cm$^{-1}$, and a frequency difference Δ ~ 19.9 cm$^{-1}$ acquired from various samples shown in Fig. 1d-g seems in good agreement with those reported for monolayer $MoS_2$,[39] thus indicating monolayer $MoS_2$ formation on Au foils at various synthesis temperatures.

It is expected that, in the sample growth zone, the concentration of $MoO_{3-x}$ will decrease with the increase of the distance between $MoO_3$ precursor and Au substrate ($D_{ss}$), leading to different coverage of $MoS_2$. In order to confirm this, a series of Au foils were successively placed on the downstream region of the quartz tube (Fig. 2a). It is found that with different $D_{ss}$ (10.0 cm, 11.0 cm and 12.0 cm) (Fig. 2b-d, ~ 530 ℃ growth), the coverage of $MoS_2$ on Au foils can effectively tuned from ~ 70%, 50% to 10%, as clearly imaged by SEM with domain size variable from ~ 500 nm to ~ 200 nm. Upon increasing the growth temperature to 680 ℃, $MoS_2$ samples with the coverage of ~ 90%, 80% and 10% can be observed with domain size tunable from 50 μm to 20 μm (Fig. 2e-g, ~ 680 ℃ growth). In this regard, though an appropriate control of $D_{ss}$ and growth temperature, a complete monolayer $MoS_2$ film can be obtained directly on Au foils *via* a facile LPCVD method. $MoS_2$ grown on Au foils may serve as a perfect candidate in diverse fields such as effective photocatalysts,[40] solar energy funnels,[41] and integrated circuits.[42]

The LPCVD monolayer $MoS_2$ on Au foils was considered to be a more achievable system for exploring the electrocatalytic HER activity than that of $MoS_2$/Au(111) prepared under UHV conditions.[38] A series of 530 ℃ grown $MoS_2$ samples with coverage from 10% to 80% (obtained in the same growth batch by only adjusting $D_{ss}$) was selected to serves as electrocatalysts. The representative SEM images of the samples are shown in Fig. 3a-c. It is worth noting that the individual $MoS_2$ flake usually present a triangular shape with very sharp boundaries. The average edge length of these MoS$_2$ flakes increases with the growing coverage, displaying distinctly statistical values of 0.20 ± 0.04, 0.43 ± 0.03, and 0.49 ± 0.05 μm corresponding to 10%, 40%, and 80% flake coverage, respectively.

The HER activities of as-grown MoS$_2$ on Au foils with different flake coverage are reflected in the polarization curves in Fig. 3d, where the Au foil serves as the reference. Notably, MoS$_2$ sample with ~ 80% flake coverage exhibits the lowest overpotential ($\eta$) of -0.1 V *vs.* reversible hydrogen electrode (RHE), suggesting a superior HER activity, and a significant enhancement of H$_2$ evolution ($J$ = 10 mA/cm$^2$) for the same sample occurs at a voltage as low as -0.198 V (Fig. 3d). However, the MoS$_2$ samples produced from the identical growth batch with coverage from 60%, 40%, 20%, to 10% exhibit increasing overpotential of -0.141 V, -0.161 V, -0.177 V, and -0.229 V, respectively. The trend depicted in Fig. 3e reveals that the HER activity attenuates with the decrease of MoS$_2$ flake coverage.

The overall Tafel slopes were measured to be variable from 61 to 74 mV/decade (Fig. 3f). The lowest Tafel slope (61 mV/decade) is achieved from the sample with 80% coverage, which is comparable with that of the UHV deposited MoS$_2$/Au(111) of 55-60 mV/decade,[38] whilst much lower than that of CVD grown few-layer MoS$_2$ on glassy carbon (GC) electrodes (140-145 mV/decade).[35] It is known that Tafel slope is an inherent property of the catalyst which is determined by the rate-limiting step of the HER. Three possible principle steps have been proposed for the HER in acidic medium.[43] The first is a primary discharge step (Volmer reaction, equation (2)):

$$H_3O^+ + e^- \rightarrow H_{ads} + H_2O \quad (2)$$

Which is followed by either the electrochemical desorption step (Heyrovsky reaction, equation (3)) or recombination step (Tafel reaction, equation (4)):

$$H_{ads} + H_3O^+ + e^- \rightarrow H_2 + H_2O \quad (3)$$

$$H_{ads} + H_{ads} \rightarrow H_2 \quad (4)$$

Under a specific set of conditions, once the Volmer reaction accounts for the rate-limiting step, a Tafel slope of ~ 120 mV/decade should result. In addition, Tafel slope of 30 and 40 mV/decade is associated with the Heyrovsky or Tafel reaction acting as the rate-limiting step, respectively. However, to date the HER mechanism for $MoS_2$ has still not been unraveled due to the complexity of the reaction. The Tafel slope of 61 mV/decade obtained in this work is close to that of $MoS_2$ nanoislands prepared in UHV[38] and defect-rich $MoS_2$ nanoplates synthesized by chemical routes,[28] suggesting the similar surface chemistry of as-grown monolayer $MoS_2$ on Au foils.

The HER process of $MoS_2$ as-grown on Au foils is schematically shown in Fig. 4a, based on theoretical[25] and experimental results[29] that the HER activity relates closely to the edge sites of $MoS_2$ flakes and the basal surfaces are catalytically inert. The edge structure of near triangular $MoS_2$ flakes should be of Mo-terminated edge, which is reactive for HER.[44] By applying extrapolation method to the Tafel plots, exchange current density ($j_0$), another parameter for the HER rate, can be deduced for all the samples mentioned above (Fig. 4b), where the sample with 80% coverage exhibits a remarkable $j_0$ of 38.1 μA/cm$^2$, by far the highest value amongst those reported for

MoS$_2$ catalyst, which is almost 120 times larger than the value for bulk MoS$_2$ (0.32 µA/cm$^2$).[33]

Statistical relations of $j_0$ as a function of coverage and edge length per area of MoS$_2$ flakes on Au foils are presented in Fig. 4c,d and Supplementary Table S1. Intriguingly, a 10% coverage LPCVD sample exhibits an exchange current density ($j_0$) of 1.51 µA/cm$^2$, much larger than that of UHV deposited MoS$_2$/Au(111) (0.31 µA/cm$^2$) at the same coverage.[38] Moreover, $j_0$ is correlated linearly with the edge length per area of MoS$_2$, suggesting a linear increase of HER activity with the growing number of the edge sites. In brief, monolayer MoS$_2$ flakes directly grown on Au foils can act as a perfect electrocatalyst for HER, with the unique traits of a wide MoS$_2$ coverage tunability and a perfect MoS$_2$-electrode contact.

The electrochemical impedance spectroscopy (EIS) was also recorded to investigate the electrode kinetics under HER operating conditions, where the Nyquist plots (Fig. 4e) present a remarkable decrease of the charge-transfer resistance ($R_{CT}$) from 1296 Ω for the 10% coverage to 25 Ω for the ~ 80% coverage sample. It is noted that the $R_{CT}$ for the 80% coverage sample is rather low as compared with related systems, and even much lower than that of the free MoS$_2$ particles ($R_{CT}$~ 10KΩ)),[30,33] accounting for the remarkable augmentation of $j_0$ in the current system. The unblocked transfer of electron between MoS$_2$ and Au, as deduced by the obvious fluorescence quenching effect from PL spectra (as will be discussed in the next part) should also be an important factor in lending to the extraordinary HER activity of MoS$_2$/Au hybrid catalyst. Another important criterion for a good electrocatalyst lies in its high durability. Figure 4f shows the polarization curve of a 40% coverage sample, where negligible differences can be noticed between the initial and after 1000 cyclic voltammetry (CV) cycling states, indicative of the excellent electrocatalytic durability of LPCVD grown $MoS_2$ on Au foils.

This study has demonstrated that the as-grown monolayer $MoS_2$ flakes on Au foils synthesized at 530 ℃ is an excellent system for examining the HER performances of $MoS_2$. The high edge density as well as a strong electron coupling between $MoS_2$ and Au foil would contribute jointly to the advanced HER activity. Note that, the polarization curves and corresponding Tafel plots of the samples synthesized at elevated growth temperatures (610 ℃, 680 ℃, and 750 ℃) usually present retarded HER activities (Supplementary Fig. S3), probably due to the gradual reduction of the density of the active sites located at the edge of $MoS_2$ (Supplementary Fig. S4 and Table S2).

In addition to the universally existing near-triangular shapes, the $MoS_2$ synthesized at higher temperature (> 610 ℃) usually manifest more complicated shapes, e.g., stars or butterfly-like shapes (Supplementary Fig. S5). Raman mapping was employed to probe the thickness uniformity and the crystal quality of monolayer $MoS_2$ flakes synthesized at higher growth temperature, with the flakes initially identified by SEM and OM images (Fig. 5a-f). Both butterfly-like individual flake and nearly complete $MoS_2$ film show quite homogeneous contrasts in the mapping micrographs (Fig. 5c,f), indicating a rather high thickness uniformity of the as-grown $MoS_2$ samples. This fact was further confirmed by single point Raman spectra randomly collected on a 20 μm × 20 μm mapping area (Fig. 5g), which display almost no shift of the typical vibrations modes ($A_{1g}, E_{2g}^1$) among different locations, consistent with that of the published results for monolayer $MoS_2$.[45] Moreover, the photoluminescence (PL) spectra of the as-grown $MoS_2$ sample was also obtained to show almost no feature at the wavelength range from 500 to 800 nm, probably due to a fluorescence quenching effect of Au foil substrates. More SEM, OM and Raman mapping images of different shapes of $MoS_2$ domain on Au foils are shown in Supplementary Fig. S6,7.

Transference of as-grown LPCVD $MoS_2$ samples onto arbitrary substrates like $SiO_2$/Si is highly demanded for revealing the intrinsic electronic property and for engineering its applications in addition to HER. Herein, a chemical wet etching method is utilized to transfer the as-grown sample. Figure 5h-j displays the OM view, Raman mapping image, and PL mapping image of $MoS_2$ samples transferred onto $SiO_2$/Si, respectively, where the nearly intact morphologies before and after transfer suggest a successful transfer of the sample on Au foils. Note that, other effective transfer method involved with a recyclable use of Au foils is still highly needed.

It is proposed that, on the as-grown sample, a strong interface interaction exists between $MoS_2$ and Au foils, enabling a suppression of the atom vibration (related to $E_{2g}^1$ mode) due to a higher force constant.[46,47] Consequently, the $E_{2g}^1$ modes are supposed to soften (red-shift). After sample transfer, Raman spectra (Fig. 5k) presents a frequency difference of Δ ~ 18.6 cm$^{-1}$ between $E_{2g}^1$ and $A_{1g}$ modes (Δ ~ 19.9cm$^{-1}$ before transfer). The $E_{2g}^1$ mode of the transferred $MoS_2$ is found to be blue-shifted by ~ 1.3 cm$^{-1}$ whereas the $A_{1g}$ mode remains nearly unchanged, which can be explained from a released interface interaction. Moreover, it is worth mentioning that a striking PL peak appears at ~ 667 nm (A excitonic emission) on transferred sample, along with a tiny shoulder peak at ~ 610 nm (B excitonic emission) (Fig. 5l), which could be addressed by the disappearance of the fluorescence quenching effect exerted by the Au substrate, namely the *n*-doping effect by Au.

In summary, we have demonstrated for the first time the scalable synthesis of monolayer MoS$_2$ on commercially available Au foils *via* a facile LPCVD route. The directly grown triangular MoS$_2$ flakes synthesized at a low growth temperature was convinced to be a perfect electrocatalyst for HER. A rather low Tafel slope (61 mV/decade) and so far the most high exchange current density of ~ 38.1 μA/cm$^2$ were achieved. We propose that, in addition to the abundant active edge sites of monolayer MoS$_2$ triangles, the excellent electron coupling between MoS$_2$ and Au foils should account for the extraordinary HER activity. Our work not only paves a new way for the controllable growth of high quality, uniform monolayer MoS$_2$ by introducing Au foil as substrates, but also offers special insight into MoS$_2$-based HER and the potential usage of monolayer MoS$_2$ in various application sectors.

**Methods**

**MoS$_2$ growth and transfer.** The growth of monolayer MoS$_2$ on Au foils was performed inside a multi-temperature-zone tubular furnace (Lindberg/Blue M) equipped with a 1-inch-diameter quartz tube. Sulfur powder, placed outside the hot zone, was mildly sublimated at ~ 102 ℃ with heating belts, and carried by Ar gas (50 standard cubic centimeter per minute (sccm)), to the downstream growth zone. $MoO_3$ powders (Alfa Aesar, purity 99.9%) and Au foils (Alfa Aesar, purity 99.985%, thickness ~ 25 μm) were successively placed on the downstream region of the quartz tube. The $MoO_3$ powders were heated from room temperature to ~ 530 ℃ within 30 minutes along with a heating rate of ~ 17 ℃/min.

The pressure for growth $MoS_2$ on Au foils was set at 30 Pa, and the growth time was set at 60 minutes under growth temperatures of 530 ℃, 610 ℃, 680 ℃ and 750 ℃, respectively. To transfer the as-grown $MoS_2$ films, the $MoS_2$/Au sample was firstly spin coated with polymethylmethacrylate (PMMA) and then baked at 170 ℃ for 10 minutes. The sample was then soaked in Au etchant for the removal of Au. Finally, the PMMA-supported $MoS_2$ was rinsed with deionized (DI) water and fished by an oxidized silicon wafer, then rinsed with acetone for removing the PMMA.

**Characterizations of $MoS_2$ films.** The prepared $MoS_2$ flakes were systematically characterized using optical microscopy (Olympus DX51), Raman spectroscopy (Horiba, LabRAM HR-800), SEM (Hitachi S-4800, acceleration voltage of 2 kV), XPS (Kratos, Axis Ultra, Mg $K_α$ as the excitation source), PL (Horiba, LabRAM HR-800, excitation light of 514 nm in wavelength).

**Electrochemical Measurements.** All of the electrochemical measurements were performed in a three-electrode system on an electrochemical workstation (CHI660D), using $MoS_2$ flakes on Au foils as the working electrode, a Pt foil as a counter electrode, and a saturated calomel reference electrode (SCE). All of the potentials were calibrated by a reversible hydrogen electrode (RHE). Linear sweep voltammetry with a scan rate of 5 mV/s, from +0.30 V to −0.35 V *vs*. RHE was conducted in 0.5 M $H_2SO_4$ (sparged with pure $N_2$, purity 99.999%). The Nyquist plots were obtained with frequencies ranging from 100 kHz to 0.1 Hz at an overpotential of 10 mV. The impedance data were fitted to a simplified Randles circuit to extract the series and charge-transfer resistances.

**Acknowledgements:** This work was financially supported by National Natural Science Foundation of China (Grants Nos.51222201, 51290272, 21073003, 21201012, 51121091, 51072004, 51201069) and the Ministry of Science and Technology of China (Grants Nos.2011CB921903, 2012CB921404, 2012CB933404, 2013CB932603, 2011CB933003), the Keygrant Project of Chinese Ministry of Education (Grants No. 313026) and the Program for New Century Excellent Talents in University (Grants No. NCET-10-0437).

**Author Contributions:** Y. F. Z. conceived and supervised the project. J. P. S. and D. L. M. developed and performed the synthesis experiments. J. P. S., D. L. M., Y. Z., Q.

Q. J., T. G. and J. Y. S., carried out SEM, OM, Raman, XPS and PL measurements. G. F. H. and X. Y. L. performed electrochemical measurements. J. P. S. and Y. F. Z. wrote the paper and all authors discussed and revised the final manuscript.

**Additional Information:** Supplementary information is available in the online version of the paper. Reprints and permissions information is available online at http://www.nature.com/reprints. Correspondence and requests for materials should be addressed to Y. F. Z.

**Competing Financial Interests:** The authors declare no competing financial interests.

**Figure Legends**

**Figure 1 | LPCVD synthesis of monolayer MoS$_2$ on Au foils. a,** Schematic view of the surface growth of MoS$_2$ on Au foils. **b,** Experimental setup of the LPCVD system. **c,** X-ray photoemission spectroscopy (XPS) spectra of as-grown MoS$_2$ on Au foils. **d-g,** Scanning electron microscopy (SEM) images of MoS$_2$ grown under distinct growth temperatures displaying different domain sizes, respectively. **h,** Optical microscope (OM) image of as-grown MoS$_2$ on Au foils. **i,** Photograph of the batch production of MoS$_2$ on Au foils. **j,** Raman spectra of MoS$_2$ flakes shown in **d-g** indicating the monolayer nature of the flakes.

**Figure 2 | SEM images of monolayer MoS$_2$ grown on Au foils with different precursor substrate distances ($D_{ss}$) and different temperature. a.** A schematic illustration of various MoS$_2$ samples with different coverage on Au foils (A–C) according to the distance ($D_{ss}$). **b-d,** SEM images of monolayer MoS$_2$ samples with coverage of ~ 70%, 50% and 10% synthesized at the same condition (at 530 ℃) but with different $D_{ss}$ of 10.0 cm, 11.0 cm and 12.0 cm, respectively. **e-g,** Similar MoS$_2$ samples grown at 680 ℃ with the coverage of 90%, 80% and 10%, respectively.

**Figure 3 | HER activity of nanosized as-grown MoS$_2$ on Au foils. a-c,** SEM images of monolayer MoS$_2$ flakes with different coverage (10%, 40% and 80%) the distribution of the domain sizes for each coverage is shown in the corresponding histogram (statistics based on SEM images including at least 100 flakes). **d-f,**

Coverage-dependent polarization curves, overpotential and corresponding Tafel plots of MoS$_2$/Au foils samples.

**Figure 4 | Analysis of MoS$_2$-based HER activities. a,** Schematic view illustrating the edges of monolayer MoS$_2$ functioning as the active catalytic sites for HER. **b,** Calculated exchange current density of different samples by applying an extrapolation method to the Tafel plots. **c, d,** Statistical relation of exchange current density with coverage and edge length of MoS$_2$. **e,** Nyquist plots of different samples. **f,** Durability test for the MoS$_2$/Au hybrid catalyst.

**Figure 5 | Raman and photoluminescence (PL) mapping of monolayer MoS$_2$ flakes prior to (on Au foils) and after transfer onto standard SiO$_2$/Si substrates (300 nm thick). a-f,** SEM, OM and Raman mapping images of a butterfly-like shape flake and a near complete film on Au foils with the Raman signal integrated from 350 to 450 cm$^{-1}$. **g,** Single point Raman spectra randomly collected from **f**. **h-j,** OM, Raman and PL mapping images of as-transferred MoS$_2$ on SiO$_2$/Si. **k, l,** Comparison of Raman and PL (PL intensity normalized by Raman $A_{1g}$ phonon peak at ~ 406.9 cm$^{-1}$) spectra between the as-grown and as-transferred MoS$_2$ samples.

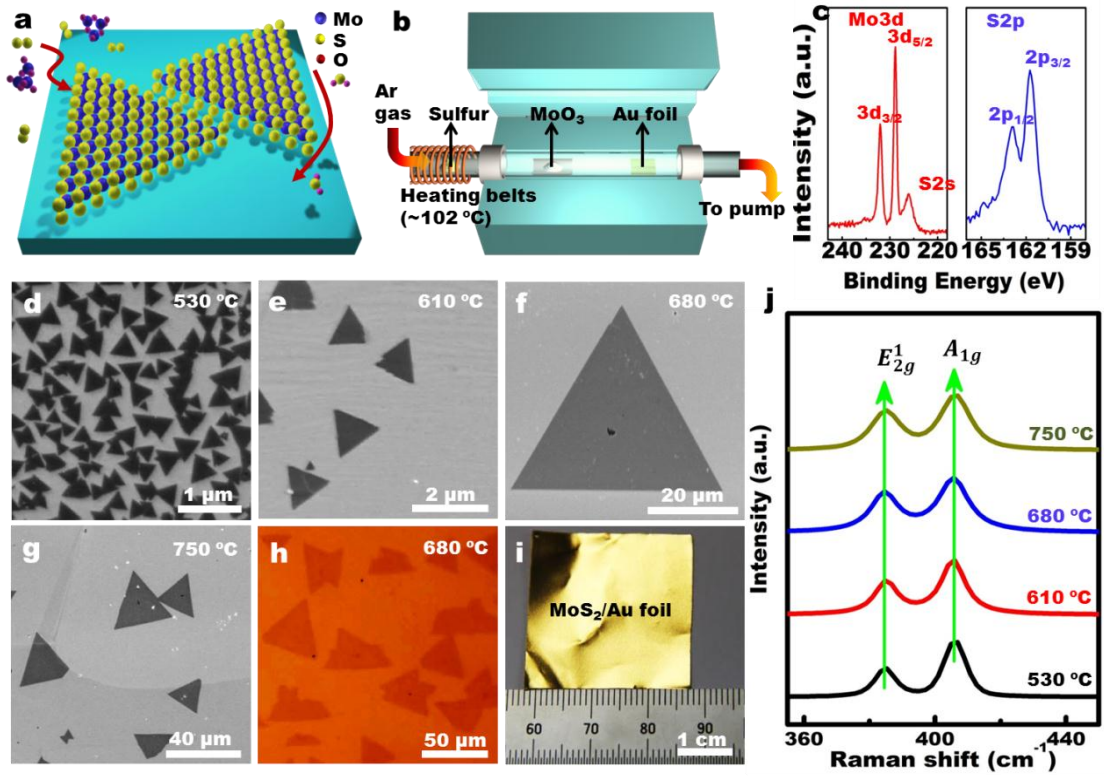

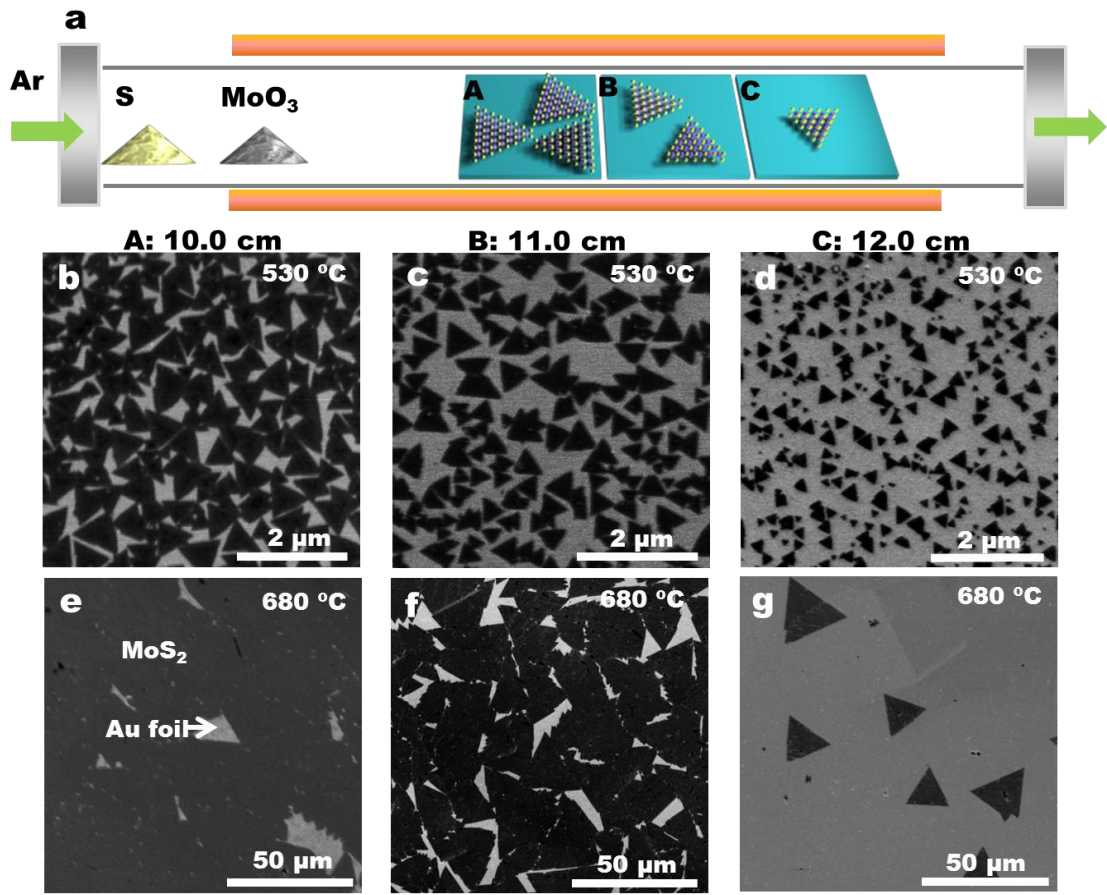

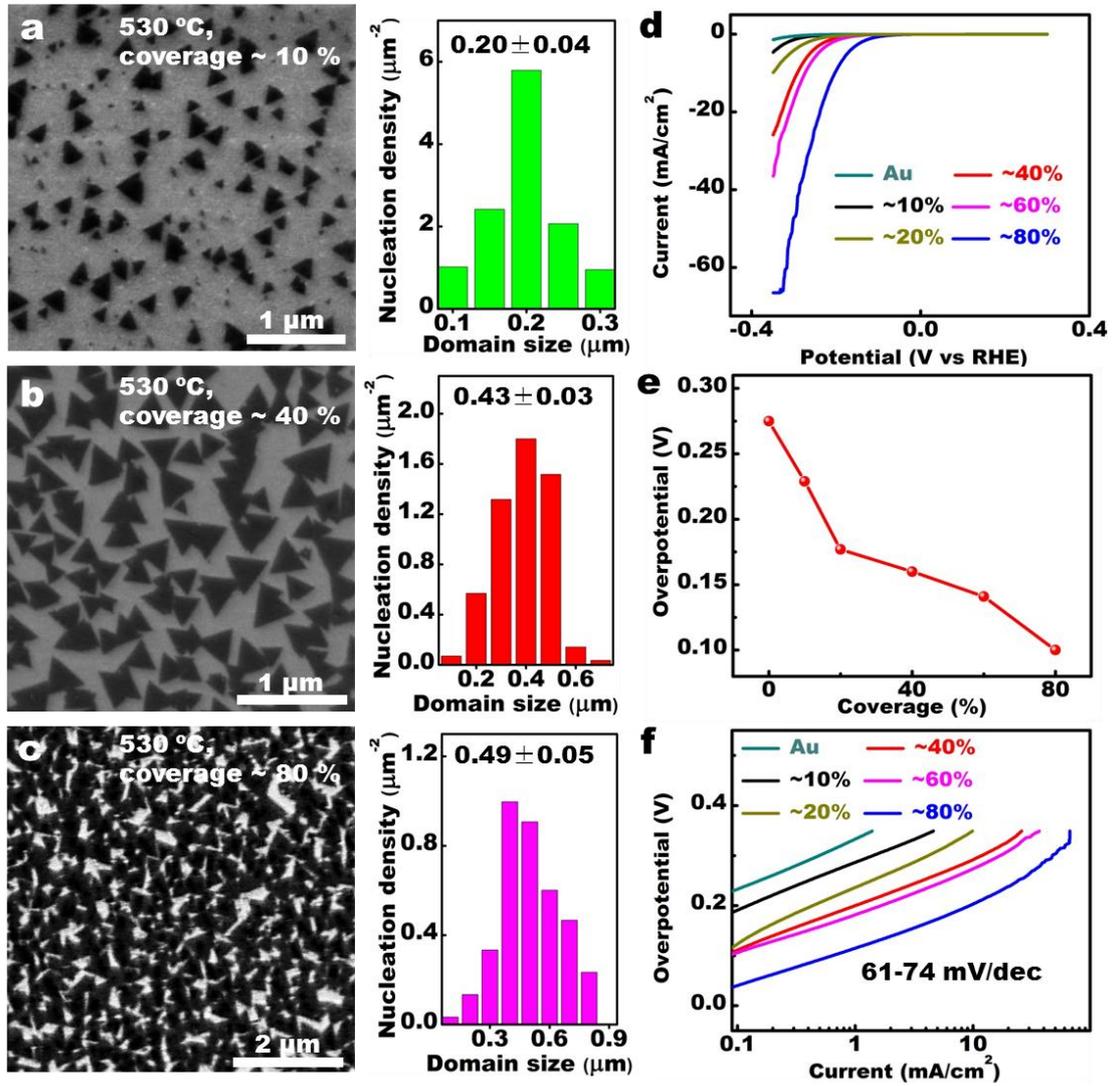

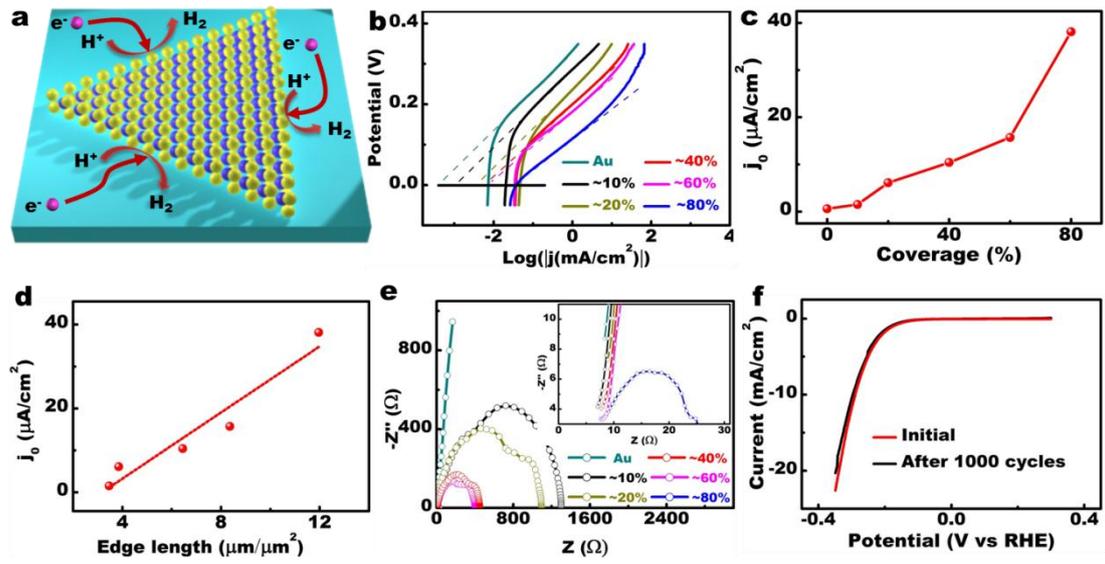

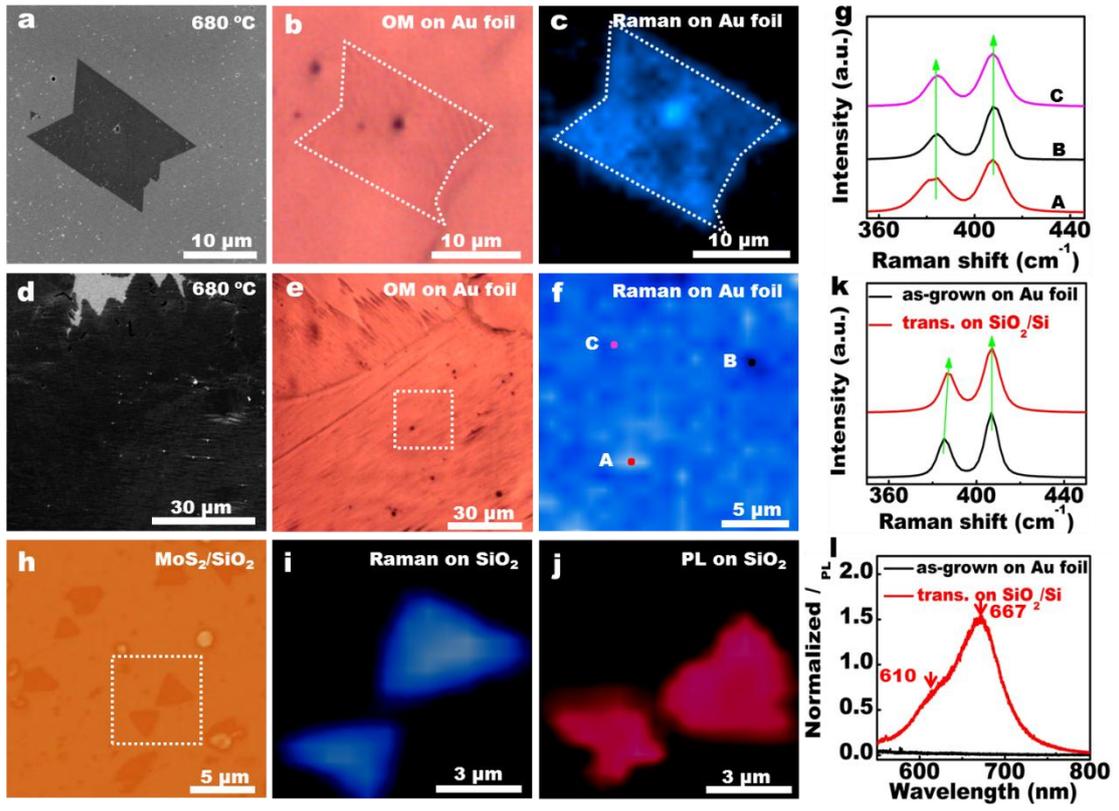